\documentclass[twocolumn]{aastex631}

\shorttitle{X-ray binary hard state from radiation GRMHD}
\shortauthors{Dexter et al.}
\graphicspath{{./}{figures/}}

\begin{document}

\title{Radiation GRMHD Simulations of the Hard State of  Black Hole X-ray Binaries\\ and the Collapse of a Hot Accretion Flow}

\author[0000-0003-3903-0373]{Jason Dexter}
\email{jason.dexter@colorado.edu}
\affiliation{Department of Astrophysical and Planetary Sciences, University of Colorado, 391 UCB, Boulder, CO 80309-0391, USA}
\affiliation{JILA, University of Colorado and National Institute of Standards and Technology, 440 UCB, Boulder, CO 80309-0440, USA}

\author{Nicolas Scepi}
\affiliation{JILA, University of Colorado and National Institute of Standards and Technology, 440 UCB, Boulder, CO 80309-0440, USA}

\author{Mitchell C. Begelman}
\affiliation{Department of Astrophysical and Planetary Sciences, University of Colorado, 391 UCB, Boulder, CO 80309-0391, USA}
\affiliation{JILA, University of Colorado and National Institute of Standards and Technology, 440 UCB, Boulder, CO 80309-0440, USA}

\begin{abstract}

We present global radiation GRMHD simulations of strongly magnetized accretion onto a spinning, stellar mass black hole at sub-Eddington rates. Using a frequency-dependent Monte Carlo procedure for Compton scattering, we self-consistently evolve a two-temperature description of the ion-electron fluid and its radiation field. For an Eddington ratio $L/L_{\rm Edd} \gtrsim 10^{-3}$, the emergent spectrum forms an apparent power law shape from thermal Comptonization up to a cutoff at $\simeq 100$ keV, characteristic of that seen in the hard spectral states of black hole X-ray binary systems. At these luminosities, the radiative efficiency is high ($\approx 24\%$) and results in a denser midplane region where magnetic fields are dynamically important. For $L/L_{\rm Edd} \sim 10^{-2}$, our hot accretion flow appears to undergo thermal runaway and collapse. Our simulations demonstrate that hot accretion flows can be radiatively efficient and provide an estimate of their maximum luminosity. 

\end{abstract}

\keywords{accretion (14) --- black hole physics (159) --- radiative transfer (1335) --- magnetohydrodynamics (1964) --- X-ray binary stars (1811)}

\section{Introduction}

Accreting black holes in X-ray binary systems (BHBs) show distinct spectral and variability states \citep[e.g.,][]{remillard2006,done2007,belloni2016}. The ``soft" states show thermally-dominated spectra peaking at a photon energy of $\simeq 1$ keV, broadly consistent with optically thick emission from a geometrically thin accretion disk \citep{shakura1973}. The spectrum in the ``hard" states is instead an inverted (hard) power law rising to a cutoff at $\simeq 100$ keV, thought to arise from inverse Compton scattering in hot, optically thin plasma \citep[e.g.,][]{sunyaev1980}. Synchrotron emission from relativistic jet electrons may also contribute \citep{markoff2001jet}.

BHB hard states are observed over a wide range of sub-Eddington luminosities, $10^{-4} \lesssim L/L_{\rm Edd} \lesssim 10^{-1}$ \citep[e.g.,][]{maccarone2003}. At the low end of this regime the infalling gas is collisionless with virial (hot) ions and colder electrons \citep[e.g.,][]{ichimaru1977,rees1982,narayan1994}. At the high end, the radiative cooling time should become shorter than the inflow time \citep{rees1982}. The resulting collapse of the hot accretion flow to a thin disk has long been associated with hard to soft state transitions \citep[e.g.,][]{esin1997}. Both the radiative efficiency and maximum accretion rate of hot accretion flows remain uncertain. Their values in analytic theory depend on the use of simplified prescriptions for angular momentum transport, electron heating, and radiative cooling \citep[e.g.,][]{yuan2014}.

It is now possible to address some of these shortcomings using MHD accretion theory. Radiative general relativistic MHD (GRMHD) simulations of weakly magnetized hot accretion flows with turbulent electron heating prescriptions have previously found low radiative efficiencies $\lesssim 1\%$ and maximum luminosities $L/L_{\rm Edd} \lesssim 10^{-4}$ \citep{ryan2017,sadowski2017}, far below that observed in the hard spectral state of BHBs. In the limit of strong magnetization and high black hole spin parameter, radiation GRMHD models of M87 have found higher radiative efficiency \citep[][]{chael2019}.

Here we perform 3D, two-temperature, radiation GRMHD simulations of strongly magnetized accretion onto a spinning stellar mass black hole with frequency-dependent Monte Carlo radiative transfer (\autoref{sec:sims}). We show that for these parameter choices, equilibrium states reach higher luminosities than previously possible, $L/L_{\rm Edd} \sim 10^{-4} - 10^{-2}$. For $L/L_{\rm Edd} \gtrsim 10^{-3}$, radiative cooling reduces the scale height of the accretion flow (\autoref{sec:results}). The emergent spectrum forms an apparent inverted power law, typical of that seen in BHB hard states. At $L/L_{\rm Edd} \sim 10^{-2}$, our hot accretion flow solution appears to undergo a cooling runaway and collapse (\autoref{sec:collapse}). We compare our results with hard state phenomenology and analytic theory in \autoref{sec:discussion}, and briefly discuss the implications for observed state transitions in BHB systems.

\begin{table*}
    \centering
    \caption{Averaged properties of the final $500 \, r_g/c$ of our \texttt{ebhlight} simulations. The mass unit $M_{\rm unit}$ and average number of superphoton packets $N_{\rm sph}$ are independent variables, while the other quantities are simulation outcomes. We italicize the M3f results since that simulation fails to achieve a steady state.}
    \label{tab:sim_table}
    \scriptsize
    \begin{tabular}{lccccccccccccc}
    \hline
Name & $M_{\rm unit}$ ($10^{10}$ g) & $N_{\rm sph}$ ($10^9$) & $\dot{m}$ ($10^{-3}$) & $\langle H \rangle / r$ & $\langle \beta \rangle$ & $\langle Q_{\rm em} \rangle$ & $\langle Q_{\rm sc} \rangle$ & $\langle \theta_e \rangle_J$ & $L/L_{\rm Edd}$ ($10^{-3}$) & $L_{\rm sc}/L_{\rm em}$ & $\Gamma_{3-20}$ & $E_{\rm peak}$ (keV) & $L/\dot{M} c^2$\\
\hline
M5 & 1 & 0.50 & 0.02 & 0.20 & 3.44 & 922 & 342 & 10.58 & 0.02 & 0.7 & 2.10 & 0.01 & 0.08\\
M4 & 10 & 1.42 & 0.44 & 0.22 & 2.57 & 2318 & 196 & 4.33 & 0.79 & 5.5 & 1.92 & 57.61 & 0.18\\
M3 & 17 & 1.48 & 1.34 & 0.15 & 0.72 & 2129 & 159 & 2.84 & 3.22 & 18.4 & 1.76 & 121.75 & 0.24\\
M3f & 30 & 1.72 & \emph{1.57} & \emph{0.13} & \emph{0.90} & \emph{2308} & \emph{122} & \emph{1.52} & \emph{6.22} & \emph{36.2} & \emph{1.66} & \emph{121.75} & \emph{0.40}\\
\hline
    \end{tabular}
\end{table*}

\begin{figure*}
    \begin{tabular}{cc}
    \includegraphics[width=0.98\columnwidth]{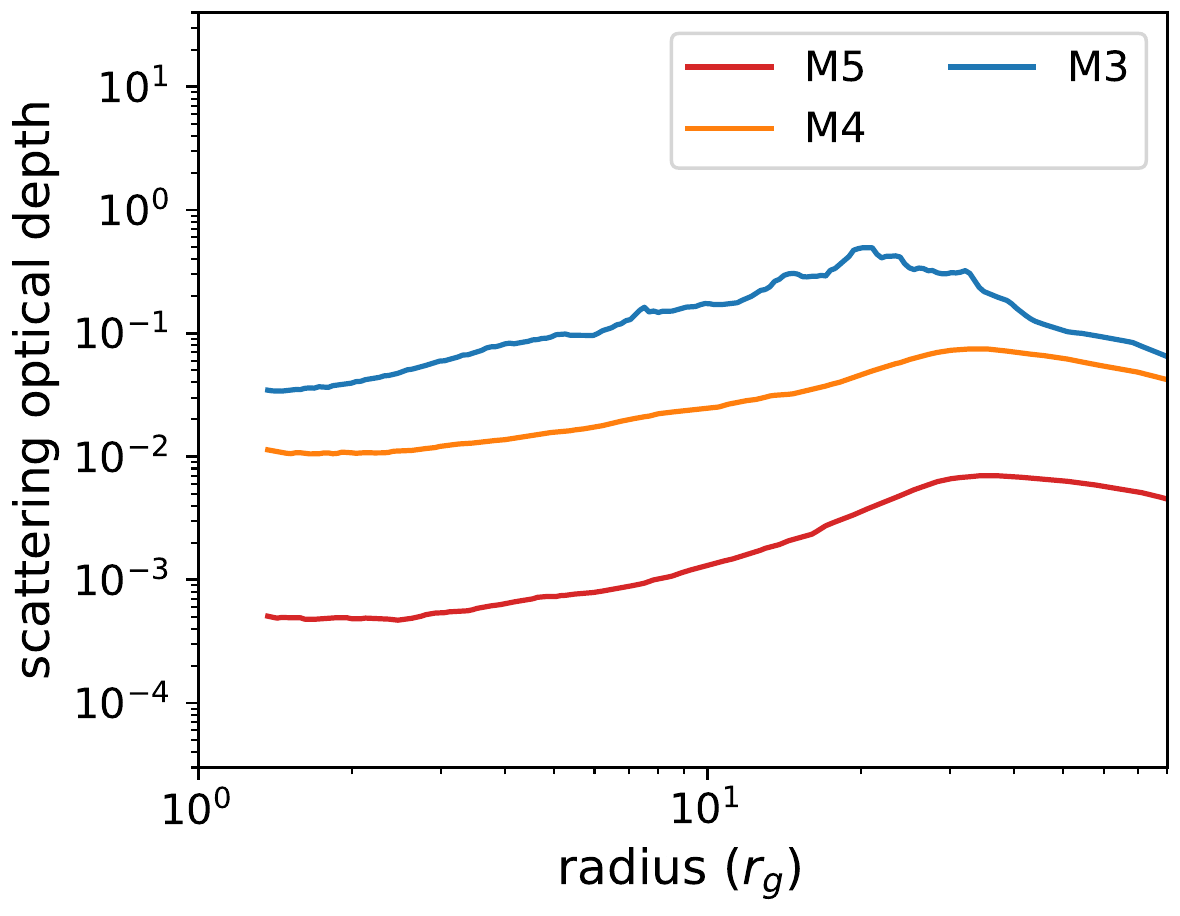} &
    \includegraphics[width=0.98\columnwidth]{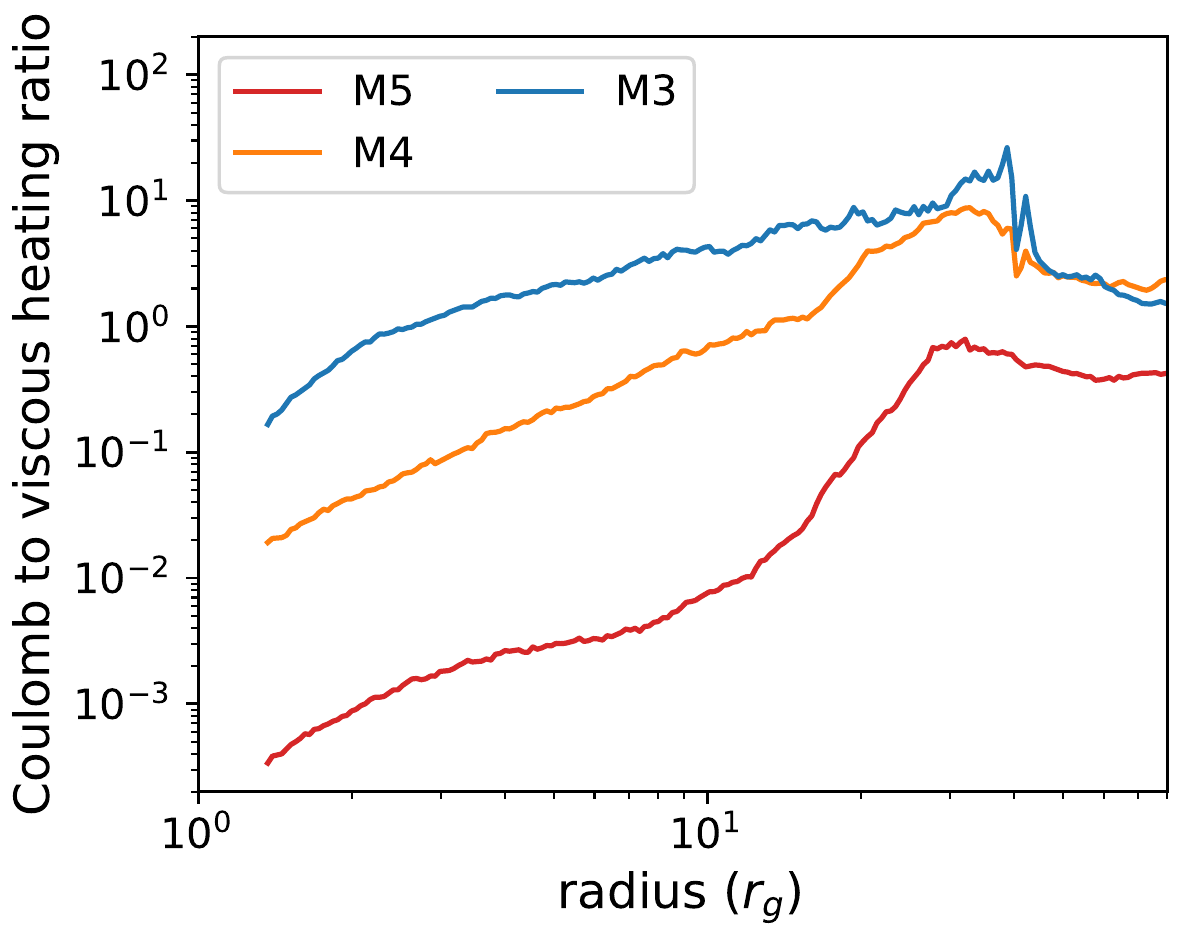}\\
    \includegraphics[width=0.98\columnwidth]{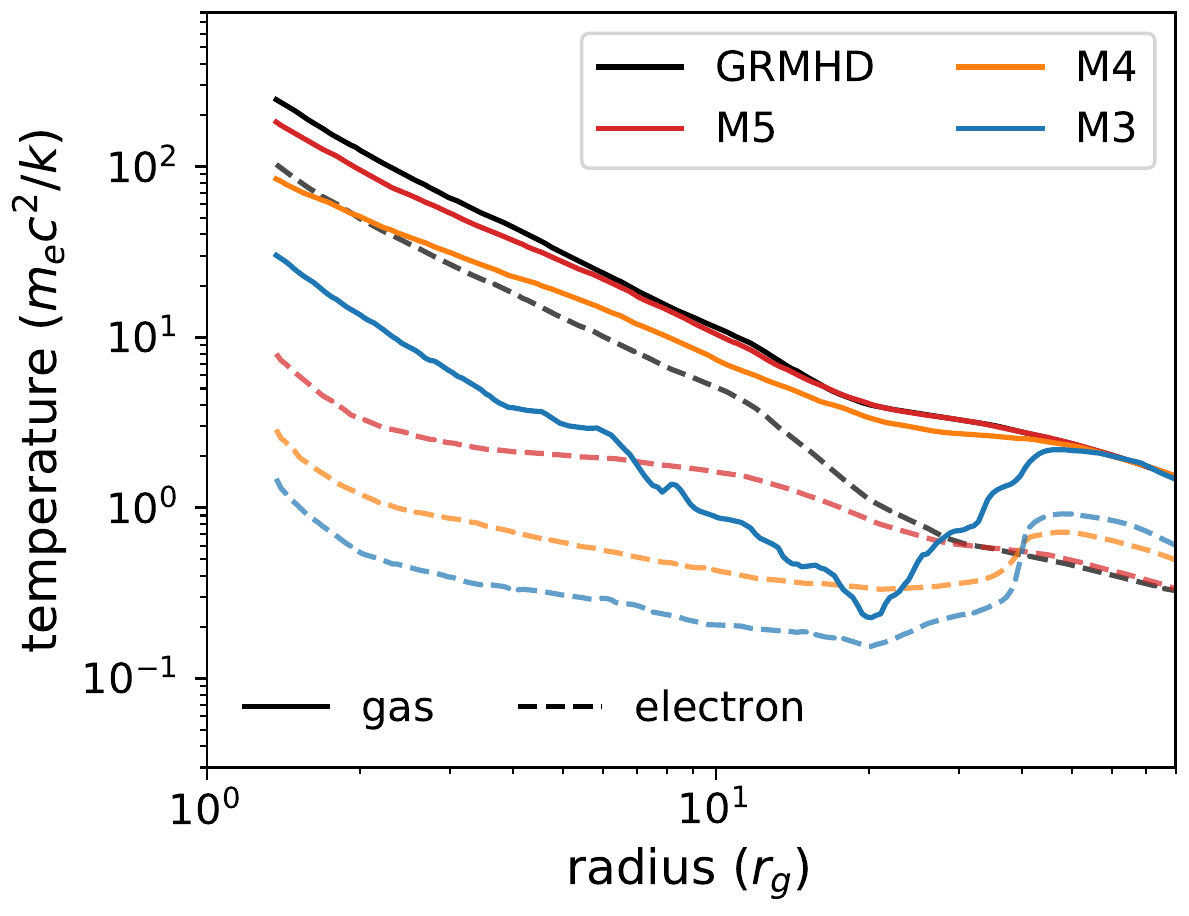} &
    \includegraphics[width=0.98\columnwidth]{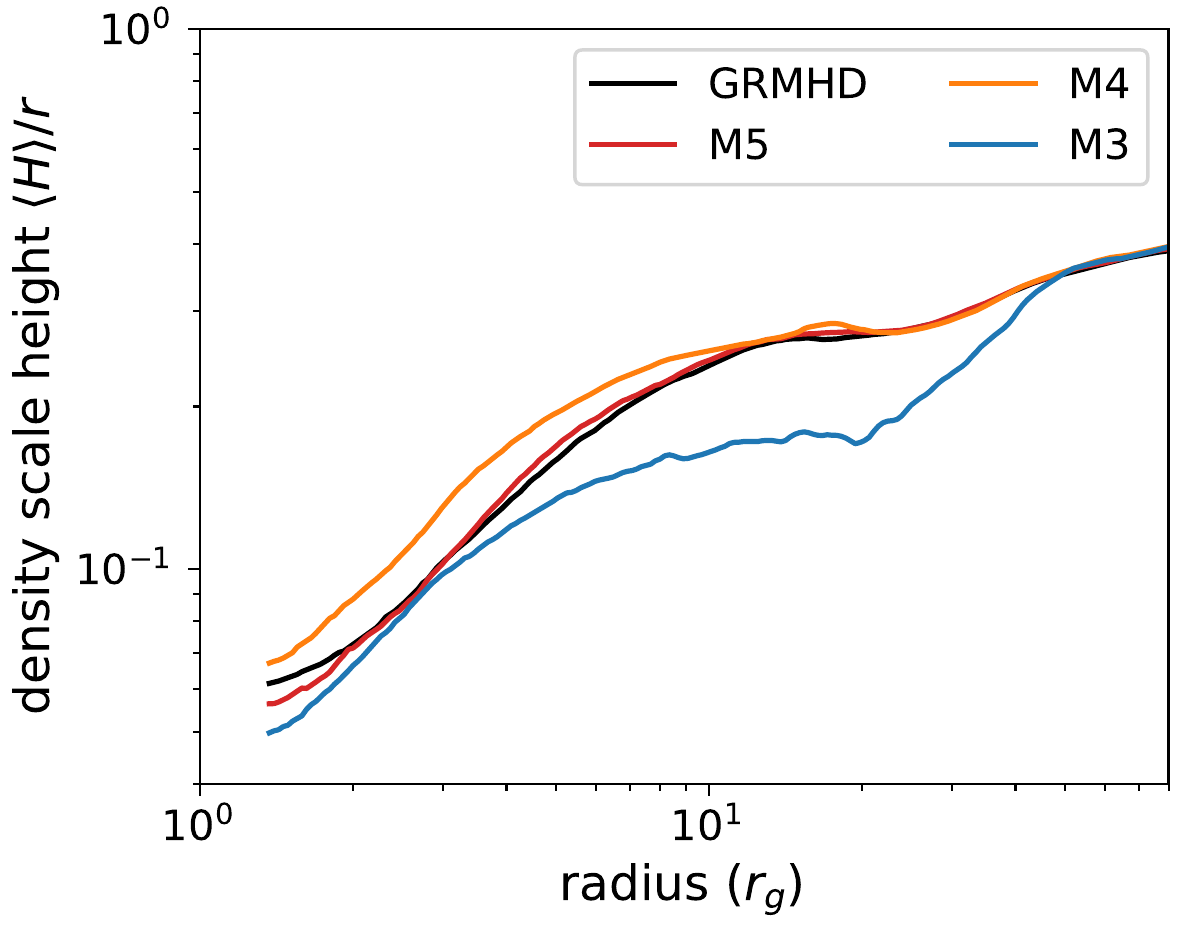}\\

    \end{tabular}
    \caption{Density-weighted, shell-averaged radial profiles for sample quantities from each simulation (colors) compared to non-radiative GRMHD (black) where possible. With increasing $\dot{m}$, the scattering optical depth increases, reaching $\tau_{\rm sc} \lesssim 1$. The disk electron temperature systematically drops to become non-relativistic, and Coulomb collisions become an important heating source for the electrons. At high $\dot{m}$, radiative cooling reduces the gas (ion+electron) temperature by a factor of $\gtrsim 10$, resulting in a decrease in gas pressure support and a reduced density scale height.}
    \label{fig:sim_plots_1d}
\end{figure*}

\begin{figure*}
\begin{tabular}{c}
    \includegraphics[width=0.97\textwidth]{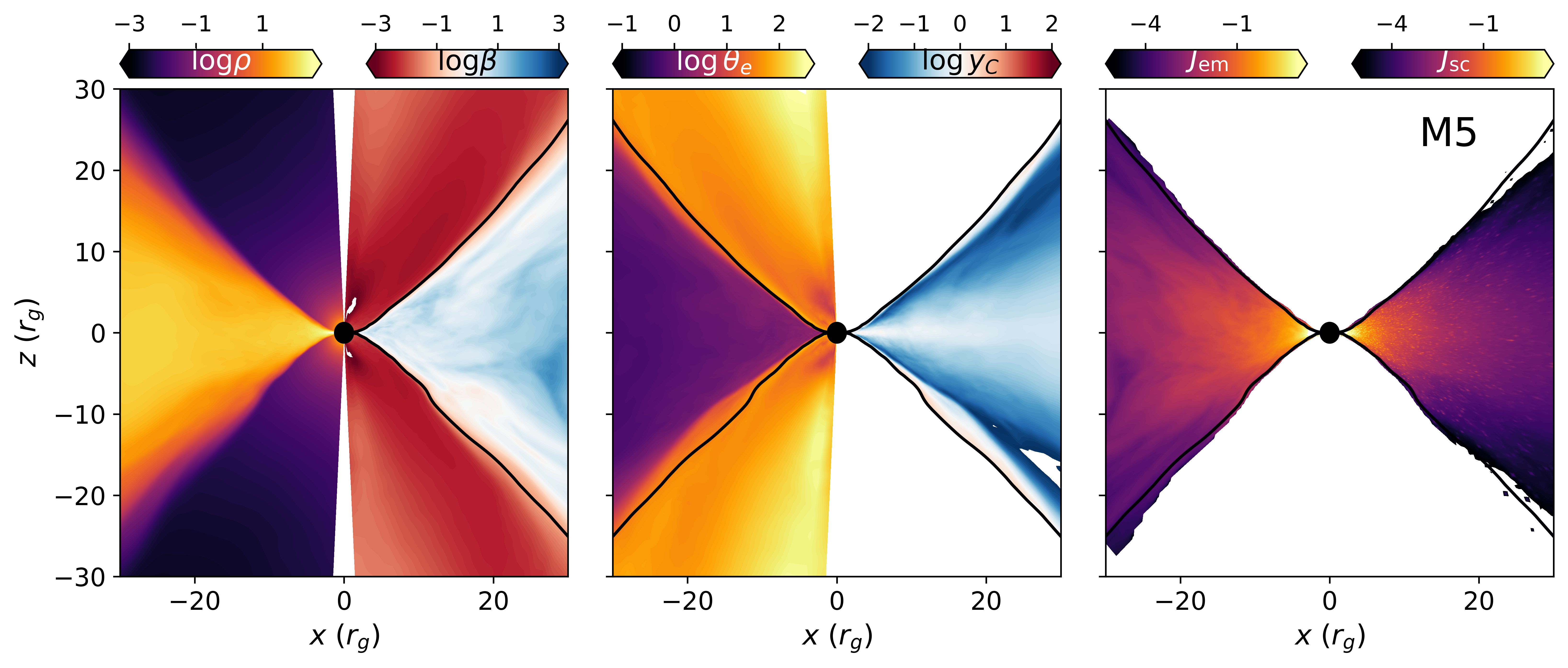}\\
    \includegraphics[width=0.97\textwidth]{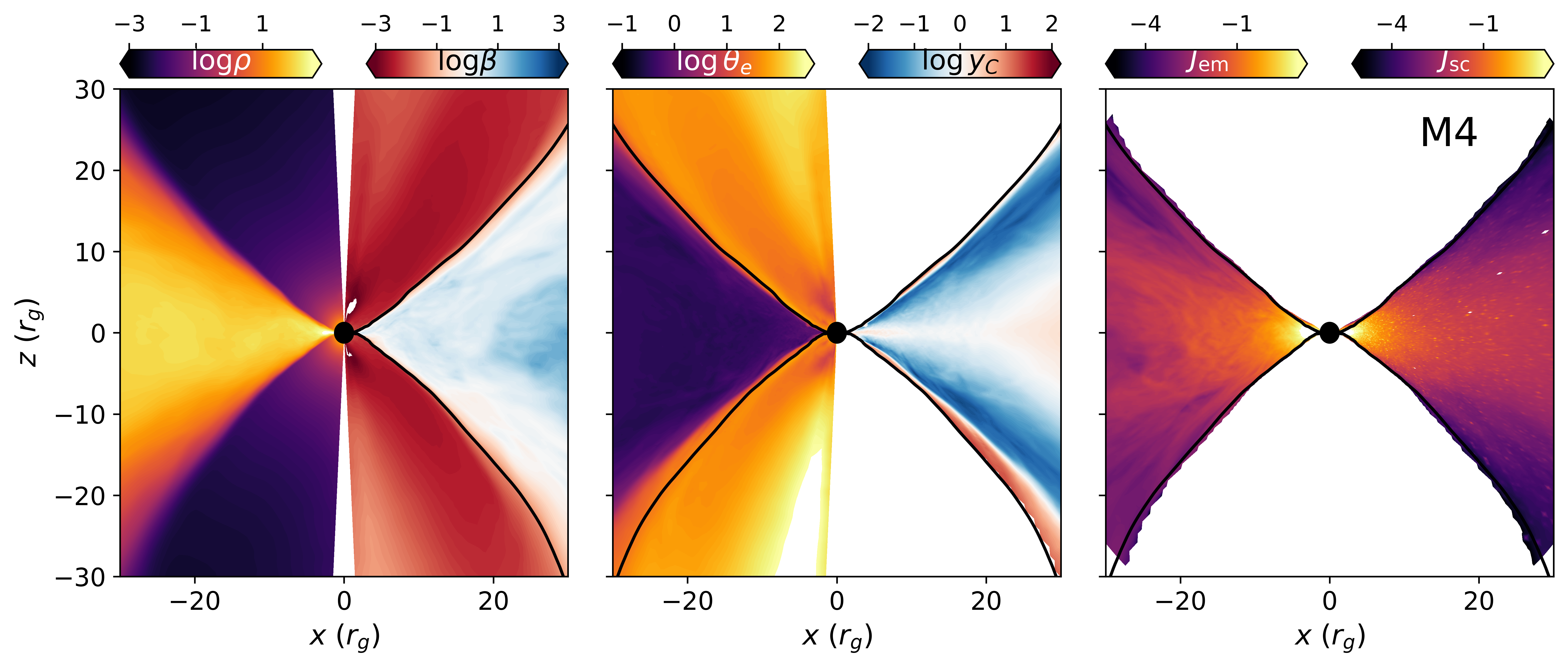}\\
        \includegraphics[width=0.97\textwidth]{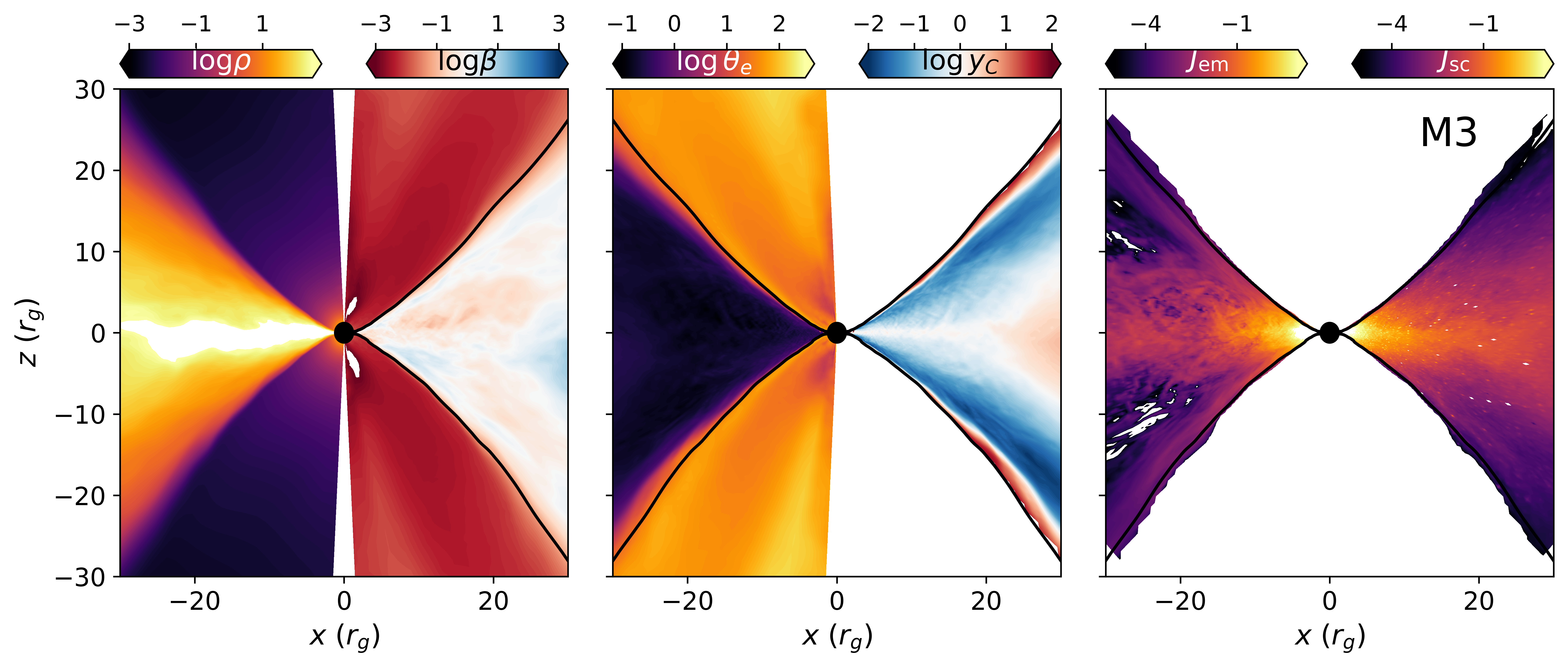}
    \end{tabular}
    \caption{Azimuthally averaged snapshots from equilibrium states of our simulations. The accretion flow Compton $y$ parameter ($y_C$, calculated using $\tau_{\rm sc}$ in the radial direction) and Compton cooling rate $J_{\rm sc}$ preferentially increase at higher $\dot{m}$, while the electron temperature $\theta_e = k T_e / m_e c^2$ decreases. Efficient radiative cooling results in the formation of a dense midplane ``disk'' in the M3 simulation, where magnetic fields are dynamically important (plasma $\beta < 1$). The black contours correspond to magnetization $\sigma = 1$.}
    \label{fig:sim_plots_azavg}
\end{figure*}

\begin{figure}
\includegraphics[width=0.98\columnwidth]{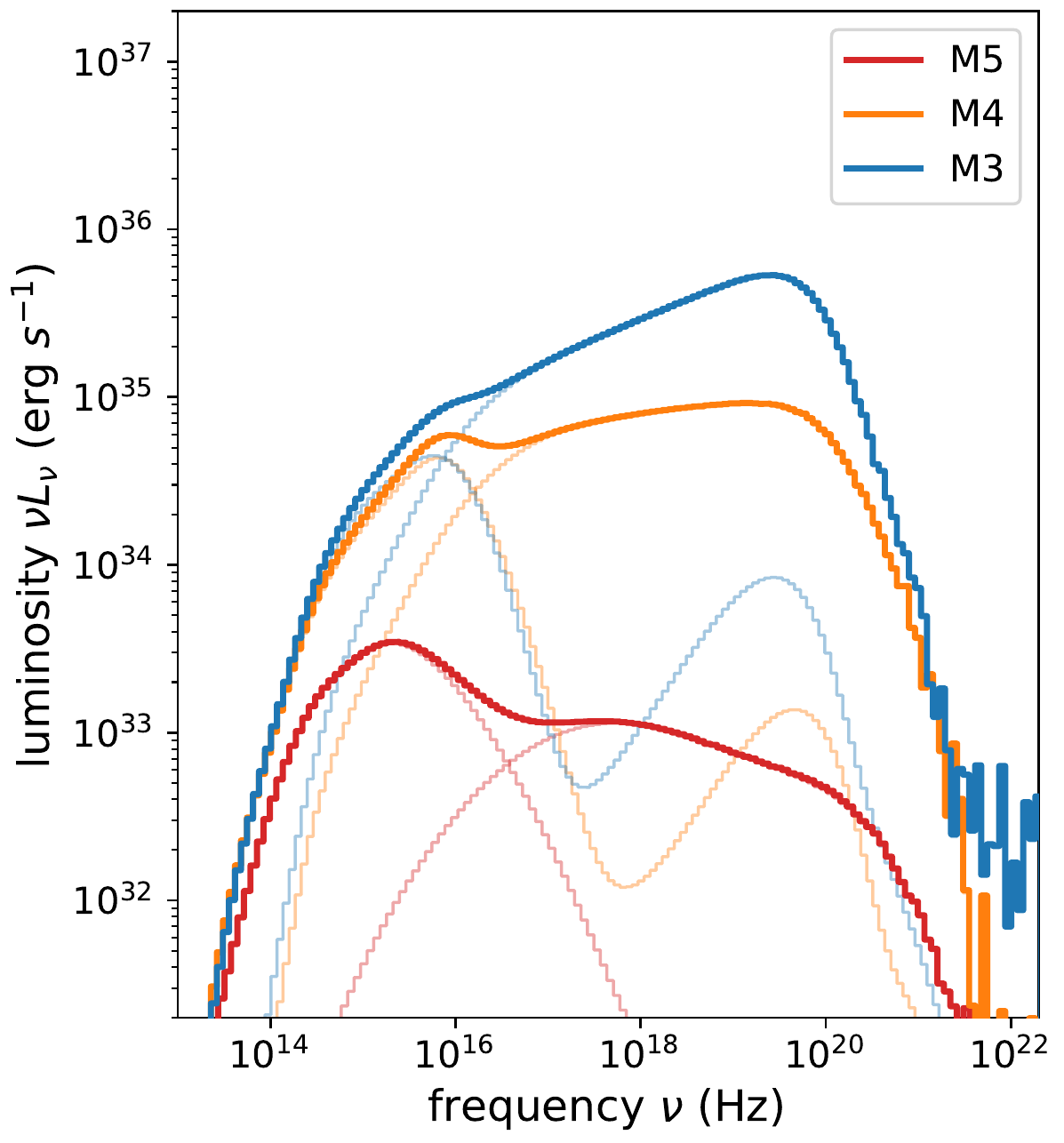}
\caption{Averaged spectrum in time and solid-angle for each simulation. Contributions from emission and inverse Compton scattering are shown as thin lines. The synchrotron spectrum peaks at $\simeq 10^{15}$ Hz with a significant X-ray inverse Compton component. Multiple scattering results in a hard power-law spectral shape for the M4 and M3 models.}
\label{fig:spectra}
\end{figure}

\section{Radiation GRMHD Simulations}
\label{sec:sims}

We carried out radiation GRMHD simulations using the public code \texttt{ebhlight}\footnote{\url{https://github.com/AFD-Illinois/ebhlight}} \citep{ryan2015,ryan2017}. \texttt{ebhlight} uses the \texttt{HARM} scheme for conservative ideal GRMHD \citep{gammie2003,noble2006} and a Monte Carlo treatment of the anisotropic, frequency-dependent radiation field. The electron entropy is evolved separately \citep{ressler2015}, with heating contributions from Coulomb collisions \citep{sadowski2017} and grid-scale dissipation. Assuming a thermal distribution, the electrons interact with photons via synchrotron and bremsstrahlung emission and self-absorption and Compton scattering.

The simulations were initialized from a gas torus \citep{fishbone1976} with an inner radius of $r_{\rm in} = 12 \, r_g$, a pressure maximum radius of $r_{\rm max} = 25 \, r_g$, and a dimensionless black hole spin parameter $a = 0.9375$, where $r_g = GM/c^2$ and we set $M = 10 M_{\rm \odot}$. The grid resolution of $320\times256\times160$  cells in modified spherical Kerr-Schild coordinates was chosen to adequately resolve both the magnetorotational instability \citep{dexter2020harmpi} and the time-dependent evolution of the mass accretion rate and magnetic flux accumulation onto the black hole \citep{dexter2020}. The torus was seeded with poloidal magnetic field, whose topology was chosen to saturate the magnetic flux on the black hole \citep[a magnetically arrested disk or MAD state,][]{bisnovatyi1974,igumenshchev2003,tchekhovskoy2011}. We adopt an electron heating prescription from particle-in-cell calculations of magnetic reconnection with a modest guide field \citep[][]{werner2018}. According to this prescription, electrons receive $f_e = 1/4-1/2$ of the total dissipated energy depending on the magnetization parameter $\sigma \equiv b^2/\rho$, where $b$ and $\rho$ are the magnetic field strength and rest mass density in $GM=c=1$ units.

We first ran a non-radiative GRMHD simulation for $t = 9 \times 10^3 \, r_g/c$. Radiation was then initialized, with a simulation mass scale $M_{\rm unit}$ chosen to approximately match target accretion rate values of $\dot{m} = \dot{M}/\dot{M}_{\rm Edd} = 10^{-5}$, $10^{-4}$, and $10^{-3}$, where $0.1 \dot{M}_{\rm Edd} c^2 = L_{\rm Edd}$. We subsequently evolved both the radiative and non-radiative simulations for an additional $(1-2)\times10^3 \, r_g/c$, sufficient to reach radiative and inflow equilibrium over the same range in radius ($\lesssim 20 \, r_g$), where the inflow equilibrium radius at time $t$ is defined by $t = r_{\rm eq}/|v^r|$ \citep[e.g.,][]{narayan2012}, and $v^r = u^r/u^t$ with $u^\mu$ the coordinate four-velocity. Superphoton packets reaching $r = 40 \, r_g$ were removed from the grid and recorded, accounting for the gravitational redshift at that location \citep{ryan2015}. No radiative processes were considered outside that radius. \autoref{tab:sim_table} lists time-averaged properties of our calculations over the final $500 \, r_g/c$. Radially-dependent quantities are averaged over the range $3-15 \, r_g$. Angle brackets denote shell-averages:

\begin{eqnarray}
\langle x \rangle &=& \frac{\int \int d\theta d\phi \sqrt{-g} \rho \, x}{\int \int d\theta d\phi \sqrt{-g} \rho},\\
\langle x \rangle_J &=& \frac{\int \int d\theta d\phi \sqrt{-g} J \, x}{\int \int d\theta d\phi \sqrt{-g} J},
\end{eqnarray}

\noindent where $\sqrt{-g}$ is the metric determinant and $J$ is the cooling rate per volume. In practice the integration is carried out in code coordinates.

The \texttt{ebhlight} radiation field is represented by a number of weighted superphoton packets, $N_{\rm sph}$, chosen so that the average number of emission and scattering events per cooling time is large, $Q \equiv \dot{N} u_e / J \gg 1$, where $\dot{N}$ is the average rate of emission or scattering events, $u_e$ is the electron internal energy density. The simulations presented here used a total of $N_{\rm sph} \simeq (5-30) \times 10^8$, or $\simeq 80-450$ superphoton packets per grid cell. The quality factors for both emission ($Q_{\rm em}$) and scattering ($Q_{\rm sc}$) interactions are $\gtrsim 100$. Such large $Q$ values should result in convergence of the radiation field and electron thermodynamics \citep{yao2021}.

We parallelized the radiation runs using 1 \texttt{MPI} process per node, and \texttt{OpenMP} across each node. The domain is split into hemispheres, each containing the full radial grid. This results in an equal number of superphoton packets on all nodes and at all times to within $\pm 5\%$ when using $64$ nodes. Our radiative simulations were $\simeq 5-30\times$ more computationally expensive than non-radiative GRMHD simulations performed with \texttt{ebhlight} at the same grid resolution. The cost increased with $\dot{m}$, due to both the higher number of superphoton packets used to resolve the radiation field and the higher rate of absorption and scattering interactions.

\section{Numerical models of the hard state}
\label{sec:results}

At low $\dot{m}$, a new radiative equilibrium is reached rapidly. The radiation is produced close to the black hole where timescales are short, and the cooling primarily modifies the electron temperature \citep[e.g.,][]{ryan2017,ryan2018,chael2019}. At higher $\dot{m}$, cooling changes the accretion flow structure and equilibrium is only reached after an inflow time from the outermost radius of interest. We determine the time range over which a new steady state has been reached by waiting for the flow to establish stationary radial profiles of fluid quantities, including the total cooling rate and gas and electron temperature. By these criteria, M3 appears to reach a stable equilibrium for $r \lesssim 20 \, r_g$ within $\lesssim 2000 \, r_g/c$. For $r > 20 \, r_g$, the gas is still cooling and inflow equilibrium has not yet been reached. We note that this procedure is complicated by turbulent fluctuations in $\dot{M}$ on comparable timescales to our time-averaging, which can produce secular changes in the normalization of $J$ and $\tau_{\rm sc}$.

\autoref{fig:sim_plots_1d} shows radial profiles of vertically integrated scattering optical depth and density-weighted shell averages of the electron heating ratio from Coulomb collisions and viscous dissipation  ($\langle Q_{\rm coul} \rangle / f_e \langle Q_{\rm visc} \rangle$), the gas and electron temperatures, and the density scale height. All runs are optically thin to electron scattering in the vertical direction, although the M3 run approaches $\tau_{\rm sc} \sim 1$ near $r \simeq20 \, r_g$. For that simulation, electron heating from Coulomb collisions exceeds that from grid-scale dissipation for $r \gtrsim 3 \, r_g$. The electron temperatures are far below those from non-radiative GRMHD: radiative cooling is important in all cases. For $\dot{m} \gtrsim 10^{-4}$, the average electron temperature is non-relativistic. In the M3 simulation, radiative cooling reduces the gas temperature and pressure and in turn the density scale height.

Azimuthally averaged snapshots of fluid and radiation quantities are shown in \autoref{fig:sim_plots_azavg} for each simulation reaching a steady state. The M5 and M4 models structurally are still geometrically thick, hot accretion flows. The high magnetization chosen here results in plasma $\beta \equiv p_{\rm gas}/p_B \gtrsim 1$ in the disk and $\ll 1$ outside of it, where $p_{\rm gas}$ and $p_B = b^2/2$ are the gas and magnetic pressure. For model M3, radiative cooling of the gas results in the formation of a strongly magnetized ($\beta < 1$), denser region close to the midplane for all radii $r \lesssim 20 \, r_g$, with a scale height $\langle H \rangle / r \approx 0.15$.

In all cases, the radiative cooling by both emission ($J_{\rm em}$) and scattering ($J_{\rm sc}$) are concentrated towards the equatorial plane in the dense accretion flow, rather than in a surrounding corona or jet. This is partly by construction: we do not allow radiation from magnetically dominated regions where $\sigma > 1$. Compton cooling becomes increasingly important at higher $\dot{m}$, as can be seen in plots of the Compton $y_C$ parameter (calculated using $\tau_{\rm sc}$ in the radial direction, right middle panels). A region with $y_C > 1$ forms in the M4 and M3  simulations near the equatorial plane and extends out to large radii.

Time-averaged spectra integrated over solid angle are shown in \autoref{fig:spectra}. The thin lines show the separate emission and inverse Compton scattering contributions. Synchrotron radiation is the dominant emission process in all cases and peaks in the optical band ($\sim 10^{15}$ Hz). A secondary peak due to optically thin bremsstrahlung grows in relative strength with increasing $\dot{m}$ (secondary hard X-ray emission peak near $\sim 10^{19}$ Hz).

For the M5 model, inverse Compton scattering produces a sub-dominant contribution to the bolometric luminosity ($L_{\rm sc}/L_{\rm em} \lesssim 1$ in \autoref{tab:sim_table}). For the M4 and M3 models, repeated inverse Compton scattering of synchrotron seed photons is the dominant cooling mechanism. The result is a hard power law X-ray spectrum extending to a thermal cutoff at a photon energy of $\simeq 100$ keV. The radiative efficiency also increases with $\dot{m}$. Its value of $\approx 24\%$ for the M3 model is somewhat higher than that of a \citet{novthorne} thin accretion disk for this spin parameter. 
\begin{figure*}
    \begin{tabular}{cc}
    \includegraphics[width=0.98\columnwidth]{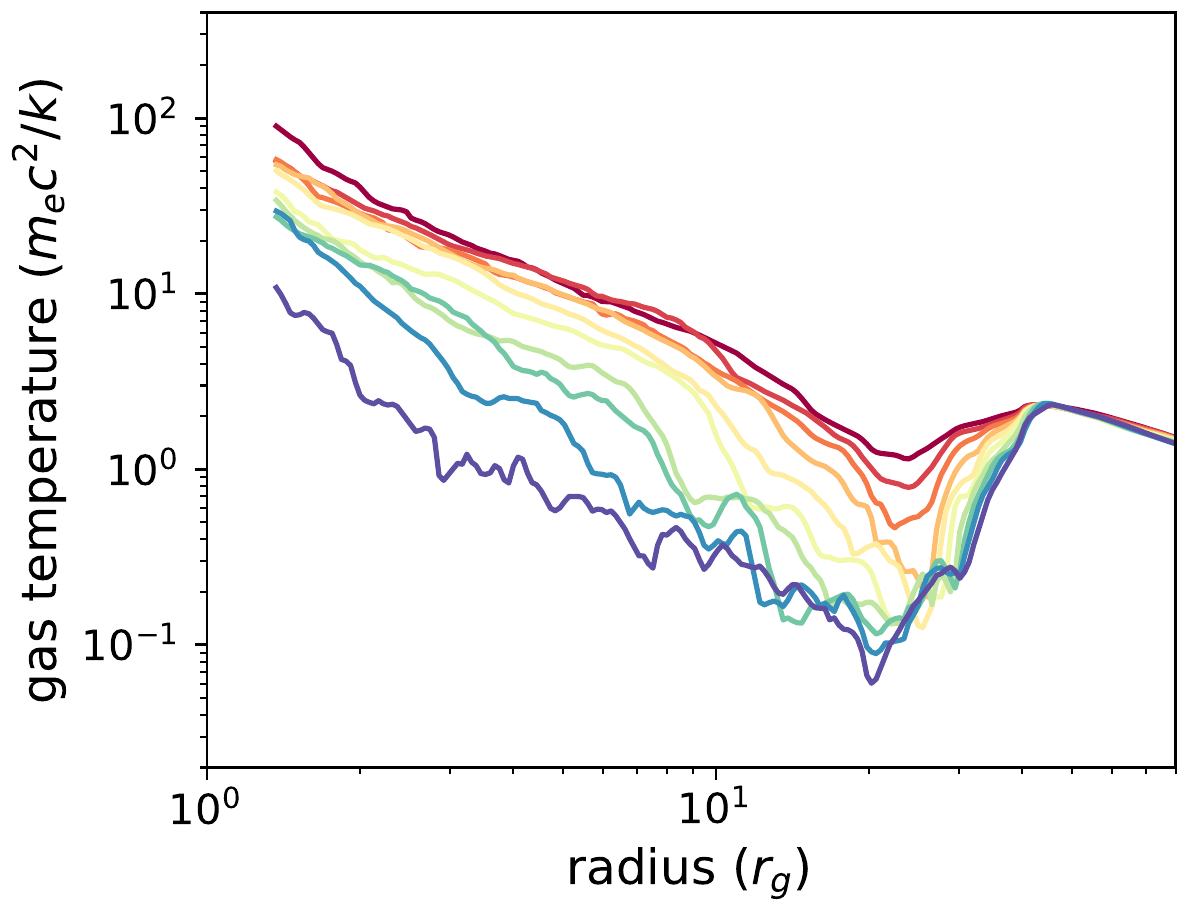} &
    \includegraphics[width=0.98\columnwidth]{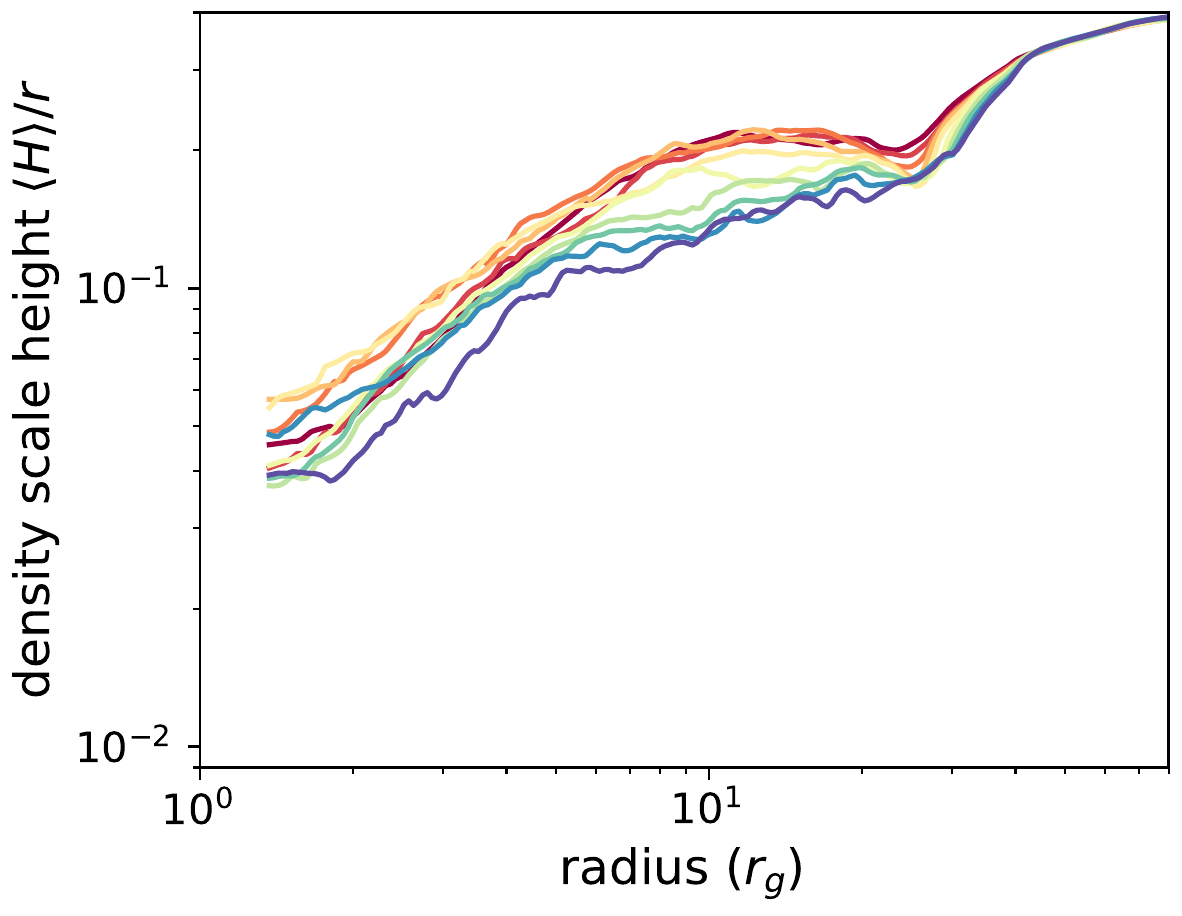}\\
    \includegraphics[width=0.98\columnwidth]{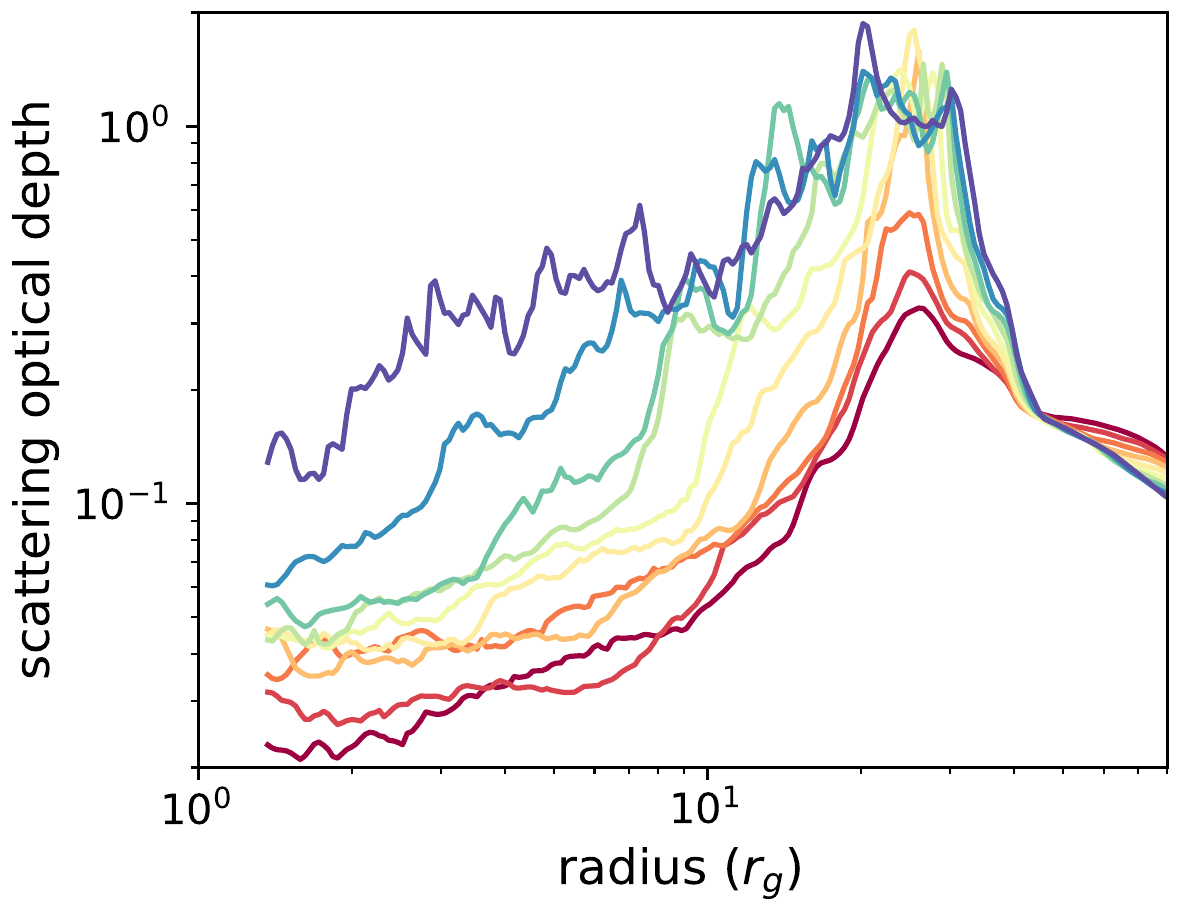} &
    \includegraphics[width=0.98\columnwidth]{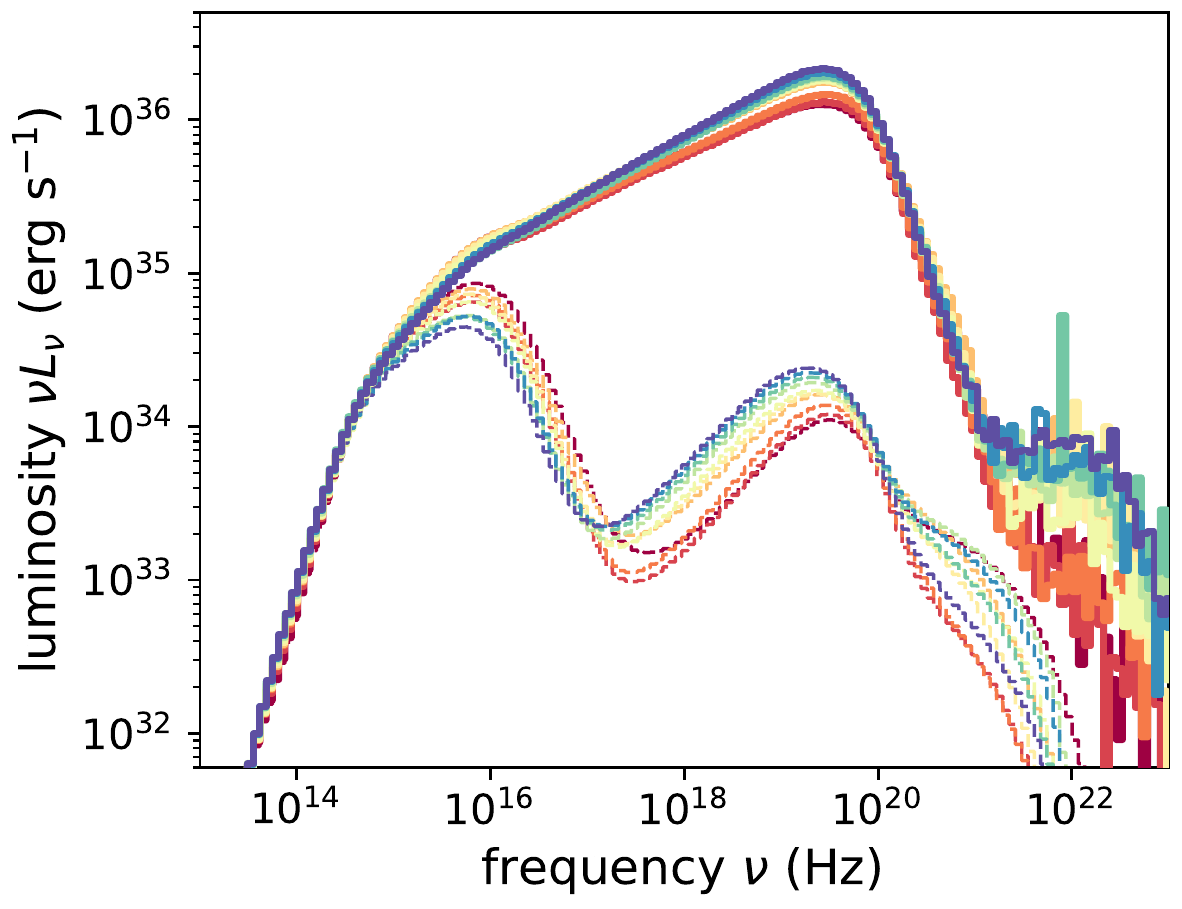}
    \end{tabular}
    \caption{Apparent collapse of the M3f simulation. The sampled density-weighted, shell-averaged profiles are spaced by $100 \, r_g/c$ over the final $1000 \, r_g/c$ from early (red) to late (blue) times. The gas temperature and density scale height steadily decrease, while the vertical scattering optical depth increases and exceeds unity at large radius. The total cooling rate and luminosity increase, and bremsstrahlung emission becomes comparable to synchrotron (dotted lines). Coulomb collisions are the main source of electron heating at all radii, and the ion-electron temperature ratio approaches $1$ for $r \gtrsim \, 5 r_g$.}
    \label{fig:sim_plots_collapse}
\end{figure*}

\section{Collapse of a hot accretion flow}
\label{sec:collapse}

For simulation M3, Coulomb coupling is a significant source of electron heating and the gas cools efficiently by inverse Compton scattering. The resulting, apparent equilibrium is poised at the edge of instability: both the scattering optical depth and Coulomb electron heating ratio are approaching unity. Once those values are exceeded, no stable two-temperature solution is expected \citep[e.g.,][]{rees1982,yuan2014}. 

We demonstrate this explicitly using a simulation with slightly higher mass unit (M3f in \autoref{tab:sim_table}). This model reached an initial radiative equilibrium at small radius $r \lesssim 10 \, r_g$. The optical depth increases to $\tau_{\rm sc} \sim 1$, and Coulomb collisions dominate the electron heating. A cooling front driven by inverse Compton scattering then propagates inwards from $r \simeq 20 \, r_g$ (\autoref{fig:sim_plots_collapse}). The loss of pressure support is strong enough to decrease the gas density scale height, while efficient Coulomb coupling drives the ion to electron temperature ratio to unity. 

As the gas temperature decreases, the total integrated cooling rate increases. Apparently the flow is thermally unstable. We are unable to follow the further evolution past the final state shown in \autoref{fig:sim_plots_collapse}. According to analytic theory, the likely result is collapse to an optically thick and geometrically thin disk.

\section{Discussion}
\label{sec:discussion}

We have presented 3D radiation GRMHD simulations of accretion onto a $10 M_{\odot}$ black hole as a function of the mass accretion rate. Our model choices of saturated magnetic flux, high spin, and efficient electron heating result in radiatively efficient accretion flows which reach luminosities of $L/L_{\rm Edd} \lesssim 10^{-2}$ while remaining optically thin. They provide relativistic, 3D, MHD realizations of physical regimes studied extensively using 1D accretion theory. They also exhibit radiative properties seen in a wealth of BHB observations.

In the hard state, BHBs show power-law X-ray spectra which harden from a photon index\footnote{The photon index $N(E) \propto E^{-\Gamma}$ is related to a spectral index $F_\nu \propto \nu^{-\alpha}$ as $\Gamma = \alpha+1$.} $\Gamma = 2$ to $1.5$ for increasing $L/L_{\rm Edd} \sim 10^{-4}$ to $10^{-2}$ \citep[e.g.,][]{remillard2006,skipper2016}. We find time-averaged photon indices of $\Gamma = 1.93$ and $1.77$ from $3-20$ keV for models M4 and M3, respectively, in good agreement with observations. The increasing spectral hardness is expected from analytic Comptonization models \citep[e.g.,][]{sunyaev1980,haardt1993}. The spectrum is exponentially cut off at a thermal energy of $\gtrsim 100$ keV, again broadly in agreement with observations of the hard state \citep[e.g.,][]{grove1998} and without requiring any non-thermal particle acceleration.

The cooling is concentrated towards the inner radii in all cases. The average emission radius of $\langle r \rangle_J \lesssim 6 \, r_g$ is consistent with observations of a compact X-ray emitting region \citep{dai2010,uttley2014}. A small amount of additional luminosity would likely be produced from the neglected region $r > 40 \, r_g$. The radiated luminosity is produced by a small fraction of the total volume, and the electron temperature weighted by emissivity, $\langle \theta_e \rangle_J$, exceeds that weighted by density, $\langle \theta_e \rangle$ (\autoref{tab:sim_table} and \autoref{fig:sim_plots_1d}). For the M3 model, $\langle \theta_e \rangle_J \simeq 3$ while $\langle \theta_e \rangle \simeq 0.2$ at the average emission radius.

The \texttt{HARM} algorithm requires the (artificial) injection of mass and internal energy in order to limit the fluid magnetization, set here to $\sigma \le 50$. We neglect electron-photon interactions whenever $\sigma > 1$ to ensure that this material does not contribute to the radiation field. The chosen cutoff value is arbitrary. At low $\dot{m}$, $\langle \sigma \rangle_J \lesssim 1$ for the M5 model and higher $\sigma$ material would likely cause order-unity changes to the emergent spectrum and bolometric luminosity \citep[cf.,][]{chael2019}. As inverse Compton cooling from larger radii becomes more important at high $\dot{m}$, $\langle \sigma \rangle_J$ decreases and the results for the M4 and M3 models should be less sensitive to the chosen cutoff value. 

Axisymmetric radiative GRMHD simulations have been carried out using \texttt{ebhlight} for $\dot{m} = 10^{-9} - 10^{-5}$ for supermassive black holes \citep[$M = 10^8 M_\odot$ or $\approx 3-6 \times10^9 M_{\odot}$,][]{ryan2017,ryan2018}. Those calculations reached $L/L_{\rm Edd} \lesssim 10^{-4}$. Qualitatively, our results are similar. The effects of radiative cooling and Coulomb coupling become more important with increasing $\dot{m}$, and a significant fraction of the radiated luminosity originates from inverse Compton scattering at larger radii. We find radiative efficiencies a factor $\simeq 10$ higher at $\dot{m} = 10^{-5}$. This is most likely due to higher accretion flow temperatures in our models, resulting from the interplay between strong magnetization (lower plasma $\beta$) and our choice of a more uniform electron heating prescription. For example, the emergent radiation here is produced by the dense accretion flow rather than the jet wall. Similar radiative efficiencies of $\simeq 10\%$ have been found in 3D radiation GRMHD models of strongly magnetized accretion onto a rapidly spinning black hole for conditions relevant to M87 \citep[][]{chael2019,yao2021}. \citet{avara2016} further showed that the radiative efficiency of MAD accretion can significantly exceed the \citet{novthorne} value, as we find for model M3. For the accretion rate range here, our models show $\simeq 3-5\times$ higher radiative efficiencies than found in analytic models \citep{xie2012}. Our limiting value of $\dot{m} \approx 10^{-3}$ is also about an order of magnitude higher than \citet{ryan2017} estimated. We see similar differences in the Coulomb to viscous heating ratio between our models and theirs, and cooling may also be somewhat more important at fixed $\dot{m}$ for higher $M$.

For $L/L_{\rm Edd} \lesssim 10^{-2}$, our hot accretion flow model M3f appears to collapse due to thermal instability in a similar fashion as long predicted by analytic accretion theory. This maximum luminosity is lower than the highest observed from hard state BHBs ($L/L_{\rm Edd} \lesssim 10^{-1}$). However, cold accretion disks appear to be present in at least some luminous BHB hard states \citep[e.g.,][]{miller2004,kara2019} and the hard power law also softens for $L/L_{\rm Edd} \gtrsim 10^{-2}$ \citep{skipper2016}. Numerical methods which do not rely on sampling individual absorption and scattering events, possibly including moment closure methods \citep{ryan2020} or implicit Monte Carlo \citep{roth2015}, will be needed to study the equilibrium flow structure and spectrum following collapse to an optically thick accretion disk.

Our models assume that the black hole has i) saturated magnetic flux and ii) high spin. Existing measurements suggest that high spin may be common in BHBs  \citep{mcclintock2011}. If sufficient magnetic flux is available from the binary companion, it may be efficiently advected in the quiescent and hard spectral states \citep[e.g.,][]{mckinney2012} at least for smaller binary systems where a hot accretion flow is expected in the outer region \citep{begelman2004}. Future parameter surveys can explore the dependence of the maximum $\dot{m}$ and radiative efficiency for hot accretion flows on the black hole magnetization and spin. We also neglect radiative cooling for $r > 40 \, r_g$, even though the scattering optical depth and Coulomb to viscous heating ratios are found to increase outwards. The evolution of the accretion flow and the accumulation of magnetic flux could be significantly altered if the flow becomes optically thick at larger radius.

The simulations presented here are $\approx 10-30 \times$ more computationally expensive than non-radiative GRMHD, and as such are limited to short durations of $\simeq 0.1$s for a $10 M_{\odot}$ black hole. Comparisons to the rich phenomenology of timing features including low-frequency quasi-periodic oscillations and energy-dependent time lags may become possible with longer duration simulations, especially if they can reach radiative and inflow equilibrium to larger radius.

\begin{acknowledgements}
JD thanks B. Ryan and G. Wong for many useful discussions about the \texttt{ebhlight} code. We thank B. Ryan, E. Quataert, and the anonymous referee for helpful comments which improved the paper. This work was supported in part by NASA Astrophysics Theory Program grants NNX16AI40G, NNX17AK55G, and 80NSSC20K0527 and by an Alfred P. Sloan Research Fellowship (JD). The calculations presented here were carried out using resources supported by the NASA High-End Computing (HEC) Program through the NASA Advanced Supercomputing (NAS) Division at Ames Research Center. 
\end{acknowledgements}

\bibliographystyle{aasjournal}

\end{document}